# Anomalous pressure dependence of the electronic transport and anisotropy in SrIrO$_3$ films

A. G. Zaitsev[1], A. Beck[1], A. K. Jaiswal[1,2], R. Singh[2], R. Schneider[1], M. Le Tacon[1], and D. Fuchs[1,*]

[1]Karlsruhe Institute of Technology, Institute for Solid-State Physics, Karlsruhe, Germany

[2]Indian Institute of Technology Delhi, Department of Physics, New Delhi 110016 India

Iridate oxides display exotic physical properties that arise from the interplay between a large spin-orbit coupling and electron correlations. Here, we present a comprehensive study of the effects of hydrostatic pressure on the electronic transport properties of SrIrO$_3$ (SIO), a system that has recently attracted a lot of attention as potential correlated Dirac semimetal. Our investigations on untwinned thin films of SIO reveal that the electrical resistivity of this material is intrinsically anisotropic and controlled by the orthorhombic distortion of the perovskite unit cell. These effects provide another evidence for the strong coupling between the electronic and lattice degrees of freedom in this class of compounds. Upon increasing pressure, a systematic increase of the transport anisotropies is observed. The anomalous pressure-induced changes of the resistivity cannot be accounted for by the pressure dependence of the density of the electron charge carriers, as inferred from Hall effect measurements. Moreover, pressure-induced rotations of the IrO$_6$ octahedra likely occur within the distorted perovskite unit cell and affect electron mobility of this system.

[*] Corresponding author: dirk.fuchs@kit.edu

I. INTRODUCTION

The 5$d$ iridium-based transition metal oxides have attracted huge interest since the spin-orbit coupling (SOC) in these compounds is on similar energy scale to that of electron-correlation or electronic bandwidth [1,2] which may in turn favor new or exotic quantum states [3-7]. In addition, the Ruddlesden-Popper series $Sr_{n+1}Ir_nO_{3n+1}$ allows for a systematic dimensional control of physical properties [8]. The SOC results in spin-orbital mixed states which, in the case of $Sr_2IrO_4$ ($n$ = 1), lead to an upper half-filled band with pseudospin $J_{eff}$ = ½ and a lower filled band with $J_{eff}$ = 3/2. The antiferromagnetic insulating ground-state makes these compounds also interesting with respect to high temperature superconductivity [9]. In contrast, perovskite $SrIrO_3$ (SIO) ($n$ =∞) displays paramagnetic semi-metallic behavior due to an increased hybridization of Ir5$d$ and O2$p$ orbitals, with a Fermi-surface consisting of multiple light electron- and heavy hole-like sheets. The 2-6 times lighter effective mass of the electrons results in an electron-like single-type carrier transport in SIO [10-12]. In contrast to the widely studied $n$ = 1 and 2 cases of the Ruddlesden-Popper series, however, the perovskite phase of SIO is only metastable under ambient conditions, which does not allow easily the growth of that material in single-crystalline form [13]. Polycrystalline materials can be obtained using high pressure sintering [14], but the perovskite phase can also be stabilized through the epitaxial growth of heterostructures such as thin films and superlattices that have been thoroughly investigated in the last decade [10,11,15-17].

Similar to bulk SIO samples, resistivities of films lightly varies between room temperature and the zero temperature limit, even though variations of absolute resistivity exist and might be attributed to different microstructures of films synthesized by different groups [18,10,19]. The low structural symmetry and SOC lead to a small electronic bandwidth which makes SIO particularly susceptible to structural changes and allows for specific tuneability of its electronic properties by means of epitaxial strain or pressure. The electronic transport of strained SIO films indeed depends sensitively on the substrate´s lattice-mismatch and may display metallic or even insulating character [10]. Interestingly, resistivity of SIO films increases significantly with increasing compressive strain [12,18], which is rather unusual for metallic systems. A strain-state analysis in combination with tight-binding calculations indicate strain-induced band-shifts in SIO films [12] . On the other hand, hydrostatic pressure seems to only weakly affect resistivity in iridates. For instance, in $CaIrO_3$ pressure only yields a small resistivity increase around 50K [20] whereas for the monoclinic modification of $SrIrO_3$ a very small decrease (-0.05% kbar$^{-1}$) is observed [21].

The electronic properties of SIO films are not only sensitive to lattice- but also to bond-angle mismatch. For example, in-plane rotations of $IrO_6$ octahedra are suppressed in very thin SIO films grown on $SrTiO_3$ (STO) due to the structural constraints imposed upon octahedral in-plane rotations by the cubic substrate and induces a metal-to-insulator transition in films thinner than 4 nm [22].

To gain insights about the intrinsic correlations between electronic transport and structural properties of SIO, it is therefore necessary to work on fully strain-relaxed films. Furthermore, as bulk perovskite SIO crystallizes in the orthorhombic *Pbnm* (62) space group, special care must be taken to ensure the growth of untwined films. Doing so, it was found that the combination of orthorhombic distortion and SOC which introduces unusual thermal expansion of the unit cell [23], which in turn results in an anisotropic electronic transport in SIO films at ambient pressure with smallest resistivity along the $c$-axis [24]. The orthorhombicity of $ABO_3$ perovskites is generally induced by the tilt of $BO_6$ octahedra and thus directly connected to the tolerance factor $t_f = (r_A+r_O)/\sqrt{2}(r_B+r_O)$, where $r_A$, $r_B$, and $r_O$ are the radii of the $A$-site and $B$-site cations (Sr- and Ir- in this case), and of the oxygen anion, respectively [25]. For pristine SIO $t_f$ = 0.9917 which yields to an in-phase octahedral rotation $\phi$ by about 9° around the $c$-axis and an antiphase tilt $\theta$ by about 12° around the $a$- and $b$-axis [23] (Fig. 1).

The tilting angles, and therefore the electronic properties of SIO are expected to be susceptible also to hydrostatic pressure $p$. Therefore, the application of hydrostatic pressure constitutes an alternative to the more commonly used substitution-induced chemical pressure or fully strained epitaxial growth and allows continuous tuning of the systems´ properties. It thereby enables a more systematic investigation of the interplay between electronic and structural degrees of freedom [26-29]. To the best of our knowledge, there is no report on electronic transport in perovskite SIO under hydrostatic pressure. In addition, recent experiments on $CaIrO_3$ demonstrated the reduction of electron correlations on the Dirac semimetal state with pressure [20] which strongly motivates transport measurements of SIO under pressure as well.

In this work, a detailed study of the electronic transport in bulk-like, nearly strain relieved, (110) oriented SIO films is reported for $0 \leq p \leq 2.7$ GPa. The importance of sample detwinning in the study of electronic transport anisotropies is emphasized.

Interestingly, an anomalous $p$-dependence of the electronic transport is observed which results in a significant increase of the electronic anisotropy, i. e., in-plane resistance anisotropy $R_{[1-10]}/R_{[001]}$ with pressure, where $R_{[1-10]}$ and $R_{[001]}$ are the resistances along the [1-10] and [001] direction of SIO, respectively . The enhancement of the anisotropic behavior is likely caused by $p$-induced structural changes. An increase of the in-phase rotation $\phi$ and a decrease of the antiphase tilt $\theta$ of the $IrO_6$ octahedra with increasing pressure may explain the contrasting $p$-dependence of the electronic transport. We show that the pressure-induced changes of the electrical resistivity cannot be accounted for by that of the electron charge carrier density, directly estimated by Hall effect measurements, which necessarily implies pressure tuning of the electron mobility.

The paper is organized as follows. A brief summary of the experimental details is given in Section II. The resistivity and electronic anisotropies of SIO at ambient pressure are discussed in section III-A. A detailed analysis of the resistivity under hydrostatic pressure follows in section III-B.

## II. EXPERIMENTAL DETAILS

The epitaxial perovskite SIO films studied in this work were prepared by pulsed-laser deposition as described elsewhere [24]. To emphasize the importance of twinning effects films were grown simultaneously on orthorhombic (110) $DyScO_3$ (DSO) and standard cubic (001) oriented STO, i. e., as delivered from the supplier (CrysTec company). In both cases, in order to avoid epitaxial strain-induced effects to larger extend and access the intrinsic properties of SIO, films with a 50 nm thickness were grown. The film thickness and out-of-plane lattice spacing of the films were deduced from x-ray reflectivity and symmetric x-ray diffraction, see Fig. 1. Despite the different lattice mismatch for SIO films on DSO and STO, the out-of-plane lattice spacing for both films is comparable, documenting a nearly strain-relived state of both films. The orthorhombic lattice parameters of SIO were deduced from a lattice parameter refinement of asymmetric diffraction peaks. The structural analysis confirmed bulk-like orthorhombic (*Pbnm*) structure with an untwinned growth for SIO on DSO and a twinned growth on STO [24]. At room temperature and ambient pressure, the lattice parameters for SIO films on DSO are very close to the bulk values (given in parenthesis hereafter) $a$ = 5.61 (5.60) Å, $b$ = 5.59 (5.57) Å, and $c$ = 7.92 (7.89) Å. The twinned growth of SIO on STO prevented accurate determination of the lattice parameters and is likely responsible for the strong damping of the Laue oscillations of the diffraction peaks shown in Fig. 1. The epitaxial relationship of the heterostructures is sketched in Fig. 2(a). For SIO on DSO the surface normal of the SIO film is parallel to the [110] direction, whereas the two orthogonal in-plane directions [1-10] and [001] are parallel to the [1-10] and [001] direction of DSO. With respect to the bulk-like lattice parameters, this in turn ensures that the octahedral rotations of the films, sketched in Fig. 2(b), are similar to that of bulk SIO. Rotations around the [001] and [1-10] directions are in-phase ($\phi$) and antiphase ($\theta$), respectively. Viewing along these directions, see Figs. 2(c,d) visualizes the impact of $\phi$ and $\theta$ on the orbital hybridization along the [1-10] and [001] direction, respectively. For twinned films on STO, domains are mainly formed with *c*-axis parallel to the [100] and [010] of STO.

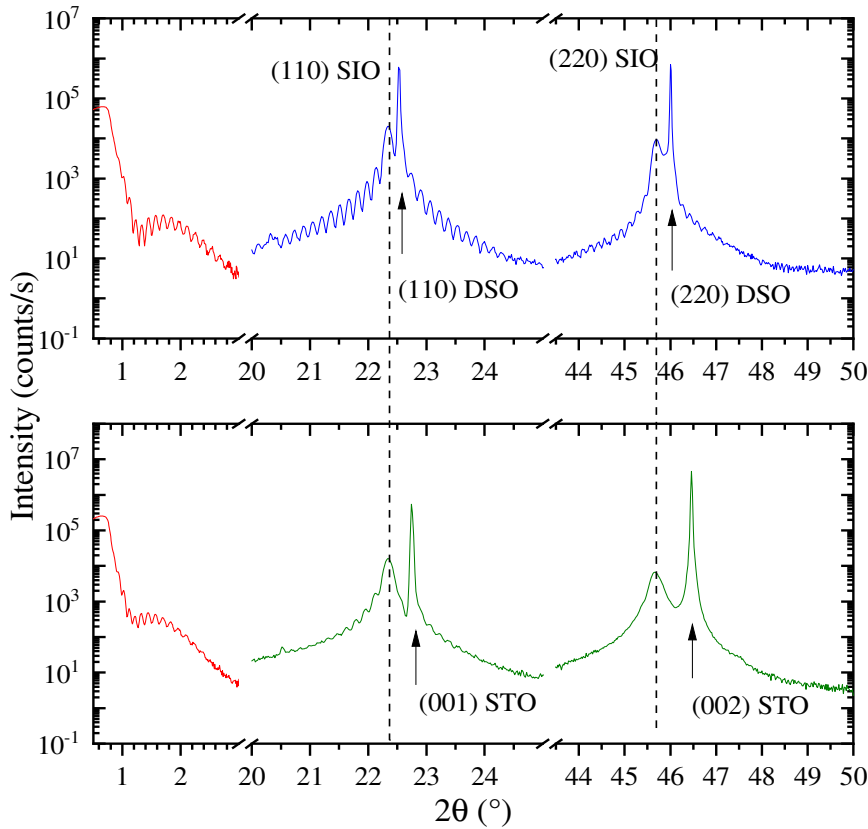

FIG. 1. X-ray reflectivity (left scan) and diffraction (middle and right scan) of a SIO film on (110) DSO (upper part) and (001) STO (lower part). The heterostructures were capped with 4 nm of STO. The SIO thickness was 50 nm for both.

To prevent possible surface degradation or decomposition, all the films were capped with a 4 nm thick epitaxial STO layer. The coated substrates were cut to 2.5 × 2.5 mm² to fit into the pressure cell. The electrical transport measurements were carried out in four-point Van-der-Pauw geometry using an Oxford He cryostat equipped with a 12-T superconducting magnet. Contacts to the buried SIO films were prepared by Al wires attached to the corners of the sample surface using an ultrasonic wire bonder. In addition, the bonds were reinforced with silver epoxy to allow an accurate positioning of the sample in the pressure cell. The pressure was applied in a two-layer clamp-type cell (C&T factory, Tokyo, Japan) which enables pressures up to 3 GPa, see Fig. 1(e). We used Daphne7373 oil (Idemitsu, Tokyo, Japan) as a pressure transmitting medium [30]. The oil was filled into a teflon cell (3.9 mm inner diameter and 17 mm length) containing the sample, ensuring that the surface normal of the substrate was aligned parallel to the cell axis within ± 2°.

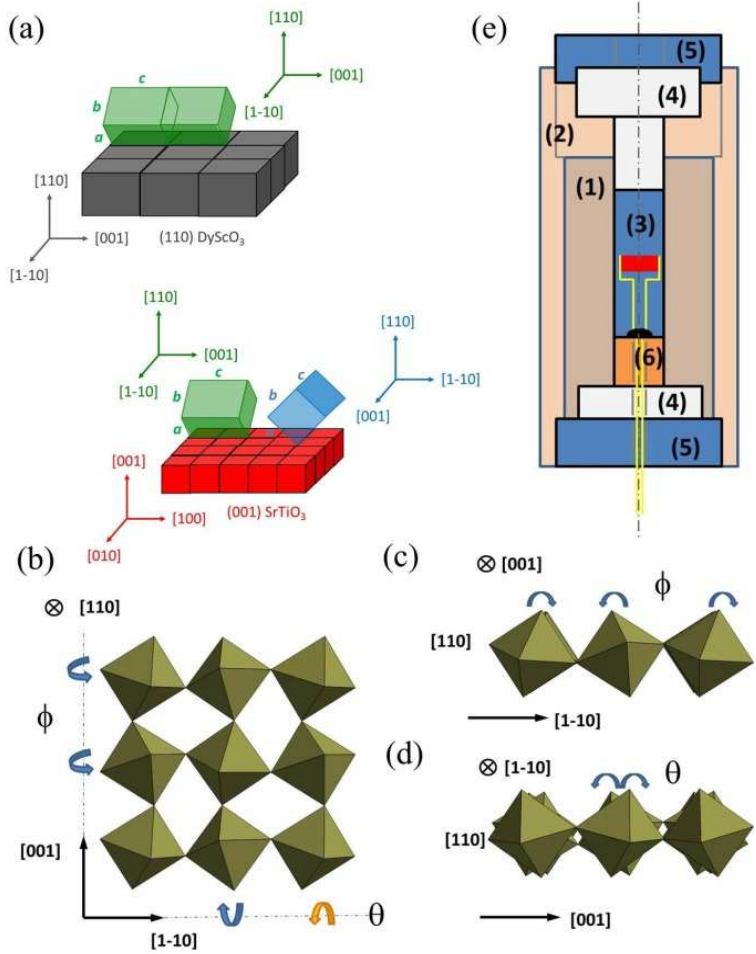

FIG. 2. (a) Epitaxial relationship of SrIrO₃ films on (110) DyScO₃ and (001) SrTiO₃. Crystallographic directions of SIO unit cells of the observed domains and the substrates are indicated. (b) Top view along the [110] direction of SIO. In-phase (ϕ) and antiphase (θ) tilt directions of the IrO₆ octahedra around the [001]- and [1-10] in-plane directions are indicated by arrows. Viewing along the [001]- (c) and the [1-10]-direction (d) visualizes the negative impact of antiphase rotation on the projected orbital hybridization along the [1-10] direction. (e) Schematic cross-sectional areal of the used two-layer clamp-type pressure cell: (1) inner NiCrAl layer, (2) outer CuBe cylinder, (3) teflon cell filled with Daphne7373 oil and loaded with wired sample (in red). Sealing was done with stycast epoxy and CuBe plug (6). (4) WC piston and backups, (5) CuBe Lock nuts.

The bulk modulus $B_0$ for SIO ($B_0$ = 187.1 GPa [31]), DSO ($B_0$ = 184 GPa [32]) and STO ($B_0$ = 184 GPa [33]) are very similar, which results in quasi-hydrostatic pressure conditions for the SIO films. The pressure was determined from resistance measurements of the superconducting transition temperature of a Pb wire [34] placed in the teflon cell. The pressure accuracy was estimated to ± 0.05 GPa. The temperature was measured via a calibrated Cernox resistance temperature sensor attached to the outside of the cell. Measurements at ambient pressure carried out before and after the application of pressure yielded the same results, showing the absence of $p$-induced irreversible changes.

## III. RESULTS AND DISCUSSION

*A. Electronic transport and anisotropy at ambient pressure*

We first focus on the electronic transport of SIO films at ambient pressure. In Figs. 3(a,b), resistivity ρ of SIO versus temperature $T$ is shown for the two orthogonal in-plane substrate directions for SIO on DSO and STO, respectively. In agreement with previous studies [10,11, 35,19], we observe in all cases a decrease of the resistivity $R(T)$ of the SIO films with increasing temperature up to a shallow minimum at temperature $T_{min}$. For the films studied here, $T_{min}$ amounts to about 200K. Considering scattering processes from impurities or various thermal excitations the Boltzmann transport equation for the quasiparticles of weakly disordered metals yield a low-temperature resistivity of the form $\rho(T) = \rho_0 + AT^q$ [36]. The second term describes classical inelastic electron scattering, which increases with increasing temperature, due to thermal excitation of phonons ($q = 1$) and increasing electron-electron collisions ($q = 2$) so that $A$ is positive. If disorder increases so that the elastic electron mean-free-path $l_e$ becomes comparable to the interatomic distance $d$, the so-called Anderson-localization of the quasiparticles occurs. The mobility edge is achieved for $l_e = d$, also known as the Mott-Ioffe-Regel condition [37,38]. Therefore, the negative resistivity slope (d$\rho$/d$T$ <0) may change sign with increasing disorder and become positive. Such a sign change has been observed frequently and studied in case of strongly disordered metals [39,40,41].

We note that resistance measurements on bulk polycrystalline SIO, also display large variabilities in $T_{min} \approx$ 10 - 60K [14,23,42,43], and in the residual resistivity (0.2 to 1.5 mΩcm). Obviously, the semimetallic behavior makes electronic transport in SIO very sensitive to disorder, imperfections, or structural distortions which may bring the system to a weakly localized state at $T > 0$.

The increase of $\rho_{1-10}$ for SIO on DSO is stronger compared to $\rho_{001}$, so that localization is expected to be more enhanced along the [1-10] direction and mainly responsible for the observed anisotropic transport. The resistivity $\rho_{1-10}$ steadily increases with decreasing $T$, whereas $\rho_{001}$ displays a shallow minimum around 220 K. The room temperature resistivity $\rho$(300K) is about 0.6 mΩcm, well comparable to values reported in the literature [10,12,44]. In contrast, SIO films on STO show similar $\rho(T)$ along the [100] and [010] substrate direction because of the twinned film growth, see Fig. 3(b).

In Fig. 3(c), we show the normalized resistance ratio $r_n$ = [$\rho_{1-10}(T)/\rho_{1-10}$(300K)]/[$\rho_{001}(T)/\rho_{001}$(300K)] for both samples. $r_n$ reflects the anisotropic electronic transport of the untwinned SIO films. In addition, for SIO on DSO $r_n$ displays a distinct $T$-dependence with a maximum around 77 K. This behavior is strongly reminiscent to that of the structural in-plane anisotropy $S = ((a^2+b^2)^{1/2}/c-1) \times 10^3$ of bulk SIO, see Fig. 3(c), that shows a maximum at the same temperature, and demonstrates the intimate relationship between anisotropic transport and orthorhombic distortion in this compound [24]. Unfortunately, the evaluation of all lattice parameters as function of $T$, which could provide more information on a distinct relationship between structural and electronic properties, remains experimentaly challenging and is out of the scope of the present study. The twinned growth of SIO on STO naturally results in $r_n \approx 1$.

In Fig. 3(d), we have plotted the mean resistivity values for the two orthogonal directions of SIO films on DSO and STO, $\langle\rho\rangle^{DSO} = (\rho_{1-10}+\rho_{001})/2$ and $\langle\rho\rangle^{STO} = (\rho_{100}+\rho_{010})/2$. The mean resistivity values are nearly the same for $T < 50$ K and differ only slightly above. This indicates, that for films on STO the distribution of SIO domains with their $c$-axis parallel to the [100] and to the [010] directions of STO is comparable. The similar behavior of $\langle\rho\rangle^{DSO}(T)$ and $\langle\rho\rangle^{STO}(T)$ further confirms that $\rho_{1-10}$ and $\rho_{001}$ are comparable for the SIO films on DSO and STO, and that the influence of additional epitaxial strain on electronic transport is essentially absent.

To understand the origin of the resistance anisotropies at a more microscopic level, we note that the Ir-O bond-lengths are the same at room temperature and differ only by about 0.5% at 3K [23]. On the other hand, the nature of the octahedral rotations in the two directions are very distinct. In the [001] direction, the IrO$_6$ octahedra are rotated in-phase, *i. e.*, only clockwise or anticlockwise around the $c$-axis, whereas rotations around the [1-10] direction are antiphase (alternating), resulting in a reduced projected orbital overlap and thus hybridization along the [1-10] direction, see Fig. 2(c,d). This indicates that the anisotropic behavior $\rho_{001} < \rho_{1-10}$ very likely originates from anisotropic orbital hybridization due to the octahedral distortions in SIO rather than from the bond-length differences which can essentially be neglected. An anisotropic orbital hybridization and electronic transport was also proposed by density functional and Boltzmann transport theory [45].

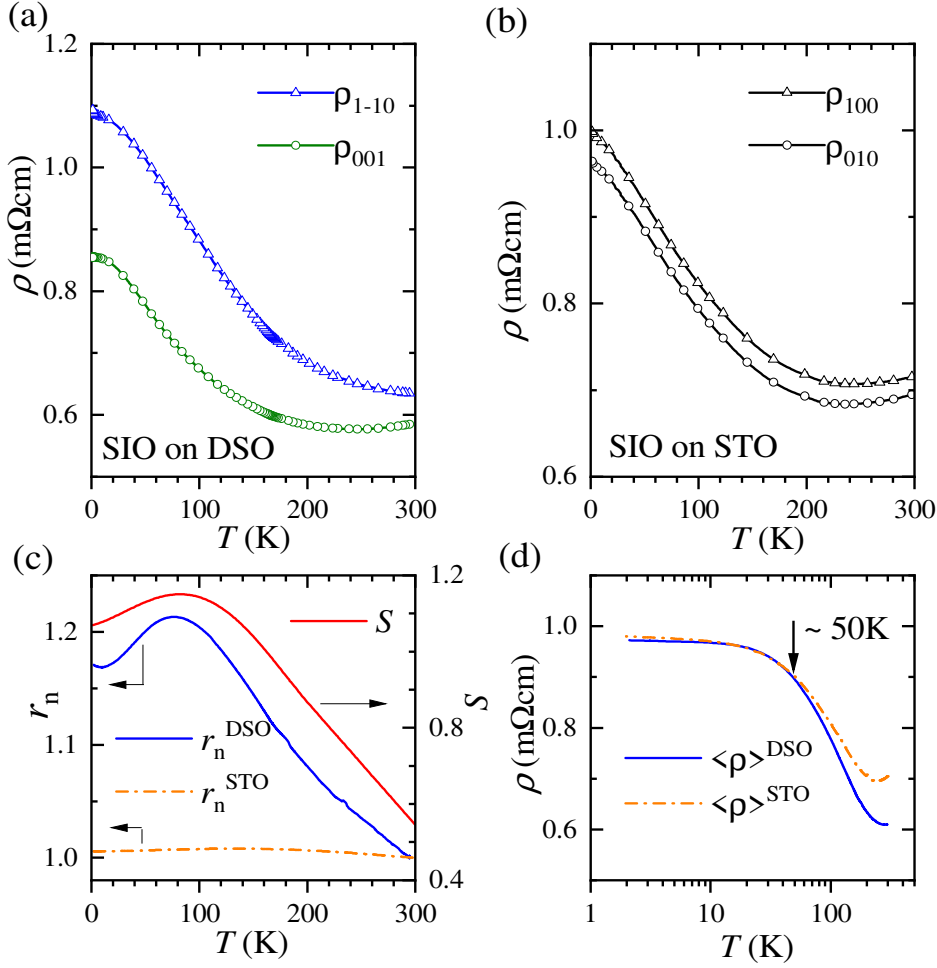

FIG. 3. Resistivity versus $T$ for the two orthogonal in-plane directions for SIO on DSO (a) and STO (b). (c) Normalized resistivity ratio $r_n = [\rho_{1\text{-}10}(T)/\rho_{1\text{-}10}(300K)]/[\rho_{001}(T)/\rho_{001}(300K)]$ versus $T$ for SIO on DSO and STO (left scale) and structural in-plane anisotropy $S = ((a^2+b^2)^{1/2}/c-1)\times 10^3$ of SIO (right scale). Data are taken from Refs.[23,24]. (d) Mean resistivity $\langle\rho\rangle^{DSO} = (\rho_{1\text{-}10}+ \rho_{001})/2$ and $\langle\rho\rangle^{STO} = (\rho_{100}+\rho_{010})/2$ versus $T$.

## B. Electronic transport under hydrostatic pressure

In Fig. 4, we have plotted $\rho_{001}(T)$ and $\rho_{1\text{-}10}(T)$ for SIO films on DSO and STO under quasi hydrostatic pressure ranging from 0 to 2.65 GPa. Interestingly, for SIO on DSO $\rho_{001}(T)$ and $\rho_{1\text{-}10}(T)$ display opposite pressure dependence. With 2.65GPa, the $c$-axis resistivity is reduced by almost 20% compared to the ambient pressure case, whereas $\rho_{1\text{-}10}(T)$ increases moderately. As expected for uniform twin domain distribution of SIO on STO, $\langle\rho\rangle^{STO}(T)$ moderately decreases with pressure.

The opposite dependence of the resistivity with pressure results in an increase of the electronic anisotropy for the untwinned SIO films on DSO. In Fig. 5(a) we have plotted $r_n(T)$ for different pressures. With respect to ambient $p$, the maximum of the resistivity anisotropy $r_n^{max}$ increases by about 10% and shifts continuously towards a lower $T_{max}$ of 63K at 2.65 GPa, see Fig. 5(b).

These results strongly suggest an increase of the structural in-plane anisotropy $(a^2+b^2)^{1/2}/c$ and a shift of its maximum to lower $T$ with increasing $p$. Therefore, $p$-induced changes of the octahedral distortion in SIO are very likely.

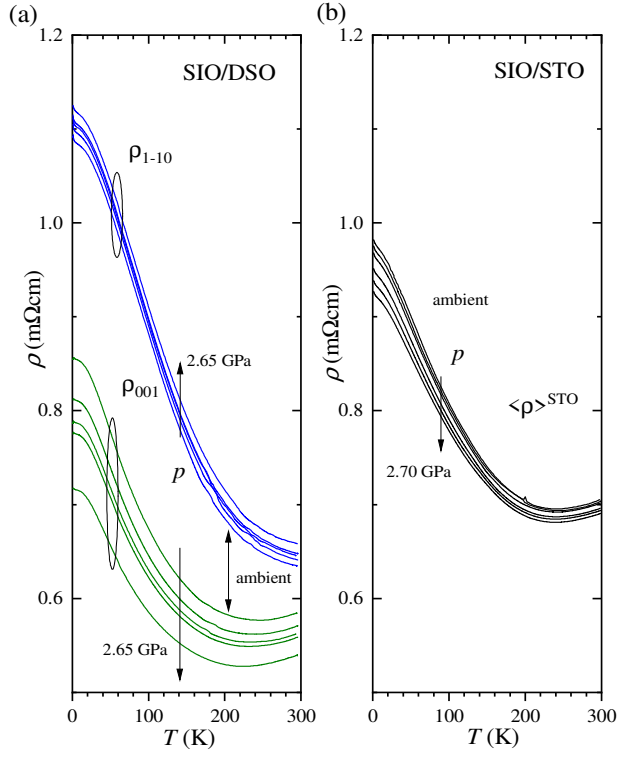

FIG. 4. (a) $\rho_{001}(T)$ and $\rho_{1\text{-}10}(T)$ for SIO on DSO for different hydrostatic pressure $p$ = ambient, 0.65, 1.35, 2, and 2.65 GPa. Data are shown by solid green and blue lines, respectively. The increase of $p$ from ambient to the maximum pressure of 2.65 GPa is indicated by arrow. Curves systematically shift with increasing $p$. (b) The mean resistivity $\langle\rho\rangle^{\text{STO}}(T)$ for different $p$. Data are shown by black solid line.

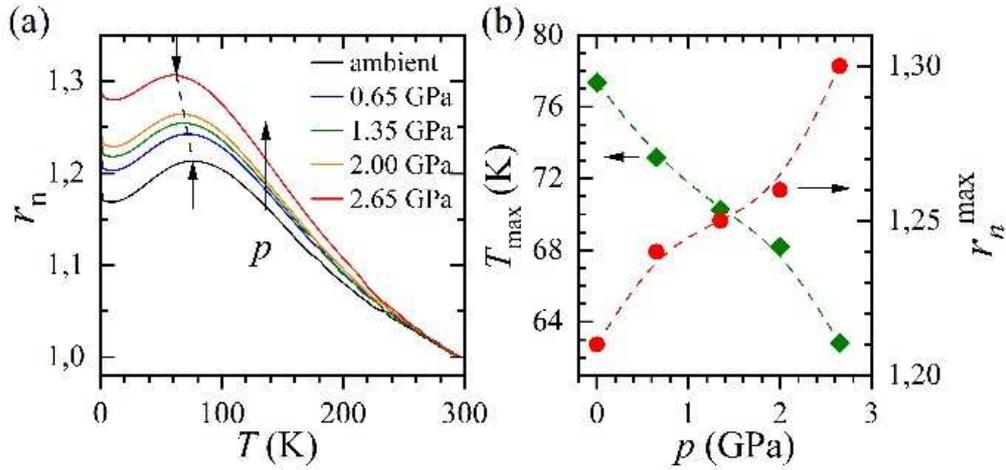

FIG. 5. (a) Normalized resistivity ratio $r_n(T)$ for different $p$. The maximum of $r_n$ is indicated by arrow. (b) The temperature, where the maximum of $r_n$ occurs (left scale) and maximum value $r_n^{\max}$ of $r_n(T)$ (right scale) versus $p$. Dashed lines are guide to the eye.

Generally, there are no strict rules that predict an increase or decrease of $BO_6$ tilt/rotation with increasing pressure for $ABO_3$ perovskites with *Pbnm* structure. Compressibility, volume and charging of $BO_6$ and $AO_{12}$ polyhedra are more specific important parameters but do not necessarily allow to deduce a general behavior for the pressure dependence of the tilt and rotation angles. In addition, for orthorhombic *Pbnm* perovskites in-phase ($\phi$) and antiphase ($\theta$) octahedral tilts are not inevitably both suppressed or enhanced by pressure [46]. For regular $BO_6$ octahedra, $\phi$ and $\theta$ can be deduced from the orthorhombic lattice parameters [47]: $\phi = \cos^{-1}(\sqrt{2}a/c)$ and $\theta = \cos^{-1}(a/b)$. Based on our discussion in section III-A, the increase of the resistivity anisotropy $r_n$ (Fig. 5a) with pressure strongly suggests an increase of the structural in-plane anisotropy, $S^*=(a^2+b^2)^{1/2}/c$. This could result from an increase of $a/c$ or $b/c$ or $a/b$ or $b/a$, which would in turn be reflected for example in an increase of $\phi$ and $\theta$, in a decrease of these two angles or a decrease of $\theta$ and an increase of $\phi$. The case of an increase of $\theta$ and a decrease of $\phi$ can be excluded. From the crystallographic point of view, all tilts are directly related to the Ir-O-Ir bending angles of the octahedral framework of the perovskite structure influencing orbital hybridization and hence conductivity along the specific directions. Because of the opposite pressure dependence of $\rho_{001}$ and $\rho_{1-10}$, the most plausible scenario would be an increase of $\phi$ and a decrease of $\theta$. An increase of $\phi$ would not affect the Ir-O-Ir bond angle and hence conductivity along the [001] direction but would reduce the Ir-O-Ir bending angle and orbital hybridization along the [1-10] direction (see Fig. 2c). Furthermore, a decrease of the antiphase rotation $\theta$ would yield an increase of the Ir-O-Ir bending angle and hence enhance the conductivity along the [001] direction while having only limited effect on the Ir-O-Ir bending along the [1-10] direction.

Possible reason for an opposite *p*-dependence of the rotation pattern, *i. e.*, $d\theta/dp < 0$ and $d\phi/dp > 0$ might be caused by different lattice contraction with increasing *p*. Similar to the increase of structural anisotropy due to the anomalous thermal expansion of SIO [23] during the decrease of *T*, one may expect anomalous lattice contraction with increasing *p* as well. As discussed by Blanchard *et al*. [23], the spin-orbit coupling in SIO which results in electronic band splitting may lead to additional lattice strain and therefore be responsible for the anomalous thermal expansion or an anomalous compressibility of SIO.

In summary, an increase of $\phi$ and a decrease of $\theta$ would be fully consistent with the observed decrease of $\rho_{001}$ and increase of $\rho_{1-10}$ with increasing pressure. The overall octahedral distortion, which can be described in orthorhombic perovskites by only one tilt $\Phi = \cos^{-1}(\cos\theta \times \cos\phi)$ around the three-fold axes of the regular octahedra [47] decreases for the SIO films on DSO with increasing pressure. For only slightly distorted orthorhombic perovskites an increase of *p* indeed usually results in a decrease of octahedral distortion whereas for strongly distorted perovskites an increase of octahedral distortion is observed [29]. However, we cannot completely rule out additional shortening of the Ir-O bond distance with increasing pressure. Nevertheless, assuming regular $IrO_6$ octahedra, this would rather lead to a similar reduction of $\rho_{001}$ and $\rho_{1-10}$ and henceforth not contribute to the anisotropy.

We finally discuss the pressure dependence of the charge carrier concentration. Hall-resistance versus magnetic field *B* (not shown here) displays perfect linear behavior over the complete range of used *T*, *B*, and *p*. In Fig. 6(a), the Hall-constant $R_H$ at $T = 2K$ is plotted as function of pressure. $R_H$ is always negative, indicating dominant electron-like transport, and its absolute value decreases with increasing pressure. Neglecting the contribution of the heavy hole-like charge carriers, we estimated the concentration of electron-like ones from $R_H$ using a single-band model, where $n_e = 1/eR_H$. At ambient pressure, the electron concentration at 2K amounts to $n_e = 3.3 \times 10^{20}$ cm$^{-3}$. The relative change $\Delta n_e/n_e$ with pressure is shown in Fig. 6(a). With respect to ambient pressure, $n_e$ steadily increases by about 6% at 2.65 GPa, which should naturally yield to an increase of the conductivity. For comparison, in Fig. 6(b) we have plotted the relative change of resistivity for both, the [001] and [1-10] direction as a function of $p$ at $T = 2$ K. The pressure-induced decrease of $\rho_{001}$ amounts to about 18%, while the increase of $\rho_{1-10}$ does not exceed 3%. The evolution with pressure of the resistivity in SIO can therefore not be accounted for only by changes of $n_e$.

This naturally leads us to suspect that the electron mobility and hence the effective mass $m_e$ are also affected by pressure. Within a simple tight-binding model, $m_e$ is inversely proportional to the electron hopping, which increases with increasing orbital hybridization. This is perfectly consistent with the scenario described above, as the pressure induced changes of $\phi$ and $\theta$ are expected to decrease (increase) $m_e$ along the [001] ([1-10]) direction. Considering the pressure dependence of $n_e$, to end up with a relative change in the resistivity $\Delta\rho/\rho$ of -18% (+3%) along the [001] ([1-10]) direction, renormalization of the effective mass $\Delta m_e/m_e$ must roughly amount to about -12% (+9%). This order of magnitude appears reasonable when compared to that of normal metals ($\Delta m_e/m_e \approx$ -6%/GPa) [4], in which hybridization generally increases with pressure.

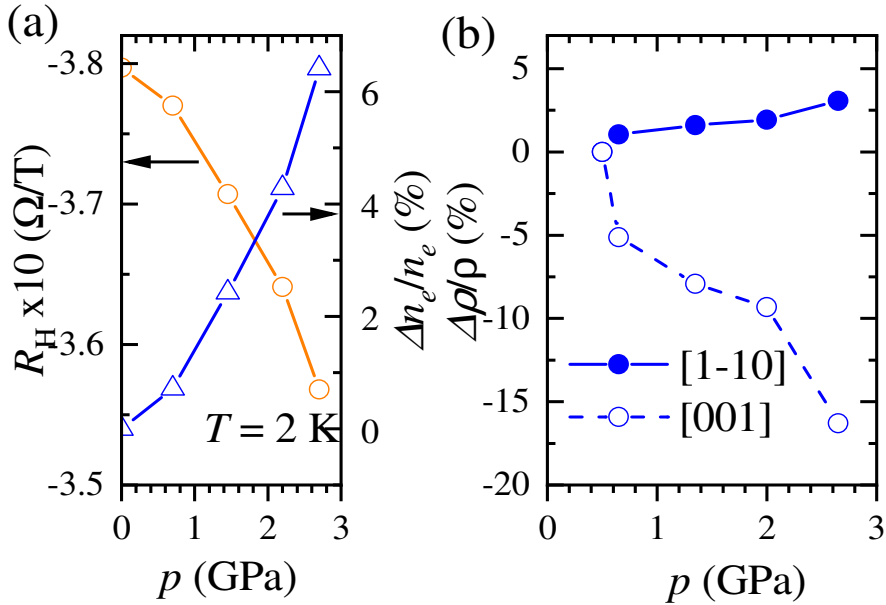

FIG. 6. (a) Hall constant $R_H$ (left scale) and relative change $\Delta n_e/n_e$ (right scale) versus $p$ at $T = 2$ K. $n_e$ was deduced from $R_H$ assuming a single-band model. At ambient pressure $n_e = 3.3 \times 10^{20}$ cm$^{-3}$. (b) Relative change of $\rho$ along the [001] and [1-10] direction at $T = 2$ K as a function of $p$.

## IV. SUMMARY AND CONCLUSIONS

The electronic transport of (110) oriented SIO films has been studied under pressure $p$. The nearly strain relieved state of the SIO films and the used DSO substrates ensure nearly hydrostatic conditions during measurements. At ambient pressure, the resistivity of untwinned SIO films slightly increases with decreasing $T$. In contrast to $\rho_{1\text{-}10}(T)$ the resistivity along the $c$-axis $\rho_{001}(T)$ displays a minimum around 200 K. The resistivity displays anisotropic behavior with $\rho_{001} < \rho_{1\text{-}10}$. The electronic anisotropy, reflected by the normalized resistivity ratio $r_n$, is very likely related to the structural in-plane anisotropy $(a^2+b^2)^{1/2}/c$ of the orthorhombic SIO which shows similar $T$-dependence.

Interestingly, $\rho_{001}$ and $\rho_{1\text{-}10}$ show opposite $p$-dependence, i. e., $d\rho_{001}/dp < 0$ and $d\rho_{1\text{-}10}/dp > 0$. At $p = 2.65$ GPa, $\rho_{001}$ decreases by $\approx 18\%$ whereas $\rho_{1\text{-}10}$ increases by 3%, resulting in an increase of the electronic anisotropy ($r_n$-1) by about 50%. The increase of $r_n$ indicates $p$-induced structural changes. In contrast to most of the perovskites where hydrostatic pressure increases or decreases both antiphase octahedral tilt $\theta$ and in-phase rotation $\phi$ simultaneously, the measurements rather suggest a decrease of $\theta$ and an increase of $\phi$. In contrast to $d\theta/dp < 0$ which increases Ir-O-Ir bending angle and hence orbital hybridization and conductivity along the [001] direction with increasing $p$, $d\phi/dp > 0$ results in opposite behavior along the [1-10] direction.

The electron-like charge carrier concentration $n_e$ increases with $p$ by only about 6% at 2.65 GPa. With respect to the changes of $\rho_{001}$ and $\rho_{1\text{-}10}$, this implies an increase (decrease) of the electron mobility along the [001] ([1-10]) direction which may be explained by $d\theta/dp < 0$ ($d\phi/dp > 0$).

## ACKNOWLEDGEMENT

We are grateful to R. Thelen and the Karlsruhe Nano Micro Facility (KNMF) for technical support with respect to atomic force microscopy. DF also acknowledges K. Sen, S. Mukherjee, and R. Eder for fruitful discussions. AKJ acknowledges DAAD for financial funding within India IIT Master Sandwich Programme (57434206).